\documentclass[11pt,a4paper]{article}
\begin{document}
\thispagestyle{empty}
\noindent
{\Large \textbf{An interpretation of the formalism of quantum\\mechanics in terms of epistemological realism}}\footnote{Paper-published in The British Journal for the Philosophy of Science \textbf{43}, 405 - 421 (1992)}
\\
\\
{\bf{Arthur Jabs}}
\bigskip
\smallskip
\newcommand{\rmi}{\mathrm{i}}
\newcommand{\rmd}{\mathrm{d}}
\newcommand{\bfitr}{\emph{\boldmath $r$}}
\newcommand{\bfitx}{\emph{\boldmath $x$}}
\newcommand{\Pal}{P_{\alpha}}
\newcommand{\ud}{\mathrm{d}}
\newcommand{\page}[1]{\hfill \mbox{{#1}\hspace{13 em}}\\}

\noindent 
\rule{\textwidth}{.3 pt}
\bigskip
\noindent
\textbf{Abstract}

\noindent
We present an alternative to the Copenhagen interpretation of the formalism of nonrelativistic quantum mechanics. The basic difference is that the new interpretation is formulated in the language of epistemological realism. It involves a change in some basic physical concepts. Elementary particles are considered as extended objects and nonlocal effects are included. The role of the new concepts in the problems of measurement and of the Einstein-Podolsky-Rosen correlations is described. Experiments to distinguish the proposed interpretation from the Copenhagen one are pointed out.
\\
\\ 
\noindent
1\enskip\   Introduction\page{} 
2\enskip\   Postulates: single particles\page{}
3\enskip\   Postulates: systems of particles\page{}
4\enskip\   The measurement problem\page{} 
5\enskip\   The Einstein-Podolsky-Rosen problem\page{}
6\enskip\   Crucial experiments\page{}
\rule{\textwidth}{.3 pt}
\\
\\
\noindent 
{\textbf{1~~Introduction}}
\smallskip

\noindent
Any physical theory consists of a mathematical formalism, that is, a set of mathematical symbols and the rules for connecting these among themselves, and a set of interpretation rules which connect the elements of the formalism with the elements of our sensuous experience. In quantum mechanics the usual interpretation rules are those of the Copenhagen interpretation. This interpretation has the peculiar feature that it rejects the language of epistemological realism because it maintains that there are properties that cannot be attributed to the objects when they are not observed but come into existence only by the act of observation or measurement. The rejection of realism was the main point in Einstein's criticism of the Copenhagen quantum mechanics. Nowadays there is a large and growing number of physicists who feel the need for a realistic formulation (Bell [1973]; Rayski [1973]; Bunge and Kalnay [1975]; L\'evy-Leblond [1976]; Bunge [1977]; Roberts [1978]; Maxwell [1982]; Burgos [1984]; Popper [1985]; Stapp [1985]; Rohrlich [1987]; Bohm \emph{et al.} [1987]; and many others).

In this paper I want to report on such a formulation. The interpretation rules of the Copenhagen interpretation are replaced by rules that are formulated in the language of epistemological realism. What is interpreted is the same as what is interpreted by the Copenhagen school, namely the standard formalism of nonrelativistic quantum mechanics. Realism, as it is meant here, is the view that the laws of nature can be formulated such as if there existed physical objects and their properties independent of whether we observe them or not. Such a realism is not a matter to be proved or disproved. It is the manner of speaking that we adopt in the normal (not philosophical) use of our language (Wittgenstein [1953]).

The main result of the present work is the demonstration that a realistic interpretation is possible, in contradiction to what is asserted by Copenhagen school. Other interpretations in terms of realism might perhaps also be possible, differing from the present one in the details of the elaboration, though to my knowledge there is none that has been elaborated to a comparable extent. A full account will be given elsewhere (Jabs [1996]). It is my opinion that the solution of the mathematical difficulties (divergences and their renormalization, e.g.), which arise with the attempt to unify quantum mechanics and relativity, will emerge only from a solution of the conceptual difficulties. Therefore, though in this paper we restrict ourselves to a reinterpretation of the nonrelativistic formalism, we hope that the concepts and ideas developed to this aim will prove fruitful in the relativistic domain as well. At least we do not know of any argument that would point to the contrary. Though we shall here argue with the help of the nonrelativistic Schr\"odinger equation this is only for convenience of presentation. We might as well have used the Dirac, Klein-Gordon. etc. equation, and use a Lorentz spinor, vector, etc. instead of the Schr\"odinger scalar $\psi$. The function $\psi(\bfitx,t)$ here actually may stand for any field in special or general theory of relativity. 

The properties which we shall attribute to the objects even when not observed are different from the conventional ones and at first sight might appear rather strange. This is the price one has to pay. It is not to be expected that the difficulties of the Copenhagen interpretation which have puzzled generations of scholars could be overcome by some cheap trick. On the other hand, once the new concepts are accepted a coherent picture emerges. Quantum mechanics is formulated in the same language as classical mechanics and any other physical theory, and the power that lies in the realistic language is available for quantum mechanics too. 

After stating our postulates in Sections 2 and 3, we shall describe how they contribute in solving the problem of measurement in Section 4, and how they account for the Einstein-Podolsky-Rosen problem in Section 5. Some crucial experiments where the realistic and the Copenhagen quantum mechanics predict different results will be described in Section 6.

\bigskip
\noindent
{\textbf{2~~Postulates: single particles}}
\smallskip

\noindent
We shall not attempt a complete axiomatization but only list the most salient features of our interpretation that are different from those of the Copenhagen interpretation. A fuller explanation with the proper motivations will be given elsewhere (Jabs [1996]).
\vspace{6pt}

(1) The function $\psi(\bfitx,t)$ (in the Schr\"odinger picture, say) is not a probability amplitude but an objective physical field, comparable to the electromagnetic field $F_{\mu\nu}(\bfitx,t)$.
\vspace{6pt} 

(2) Any particular normalized $\psi$ function, such that in the Copenhagen interpretation is the probability amplitude of the observed behaviour of an elementary particle, in our interpretation is the individual elementary particle itself. An elementary particle is defined as a particle listed in the Particle Properties Tables (Wohl \emph{et al.} [1984]). This implies that the $\psi$ function (i) is not merely a calculational device for calculating the probabilities of specified outcomes of observations, as in the Copenhagen interpretation, (ii) that $\psi$ does not merely describe `knowledge' (Heisenberg [1958]), (iii) that it does not describe an ensemble of particles, as in the statistical interpretation (Ballentine [1970]), and (iv) that it does not describe nuclei, atoms, molecules, etc., nor quarks. 

There is thus no wave-particle duality, and quantum mechanics becomes a field theory. The particle is an extended object, of size $\Delta x = \langle\,(x - \langle x\rangle)^2\,\rangle^{1/2}$, say. A single sharp position is thus not among the properties of an elementary particle. The idea that elementary particles are extended objects has repeatedly appeared in the literature. However, the size has been considered to be fixed, for example equal to the Compton wavelength $\hbar/mc$ or to the classical (electron) radius $e^2/mc^2$. In the new interpretation the size is not fixed but varies in time according to the variation of $\psi(\bfitx,t)$. For each kind of particle a specific type of a field is conceived: there is an electron field, a proton field, a photon field, etc., as it is familiar from quantum field theory. 

An elementary particle thus described is called a wavepacket or, more specifically, an elementary wavepacket. Another object which is also a wavepacket (a coalesced wavepacket) is the system of similar elementary particles that have coalesced (point (5) below). Wavepackets are the basic objects in our interpretation. The term wavepacket does, however, in no way mean a restriction to a linear superposition of plane waves, and even when it is mathematically expressed as such it does not mean that plane waves were physical constituents of the particles. It just means a special region of nonvanishing field.

This concept has much in common with Schr\"odinger's early wavepacket picture (Schr\"odinger [1926], [1928]). The wavepackets here are, however, endowed with additional properties, not encountered in ordinary quantum mechanics. The wavepacket idea may clash with some of our favoured ideas on what an elementary particle is, but the decisive counter-argument could only be a clash with observed facts. We have not detected any such clash.
\vspace{6pt}

(3) As the elementary particle is an extended object we can no longer speak of \emph{the} position of the particle, and the expression $|\psi(\bfitx,t)|^2\rmd^3x$ can no longer be the probability of observing, at time \emph{t}, the position within $\rmd^3x$ about \bfitx, as in the Copenhagen interpretation. In our interpretation $|\psi(\bfitx,t)|^2\rmd^3x$ is not defined as a probability. It is, in certain circumstances, only numerically equal to a probability, but to one that is different from the Copenhagen position probability, and may be called an action probability. It is the probability that the wavepacket $\psi(\bfitx,t)$ induces an effect (observed or not) within the volume $\rmd^3x$ about {\bfitx} and during a small but otherwise arbitrary time interval $\Delta t$ about $t$. 

This interpretation of $|\psi|^2\rmd^3x$ can be justified by a calculation that is based on the transition-probability formulas of the standard formalism (Jabs [1996]). In these formulas the probability refers only to the occurrence of the transition, not to the nature of the thing that makes the transition. And the circumstances in which this interpretation turns out to be possible include those where the physical environment of the wavepacket, in the language of conventional quantum mechanics, allows the position of the object to be directly measured. Examples are bubble chambers and photographic emulsions. The value of the time interval $\Delta t$ does not appear explicitly because the environment is meant to be homogeneous, time-independent, and such that the wavepacket with certainty induces an effect anywhere in it during $\Delta t$. Of course, $\Delta t$ must be short enough so that $|\psi|^2$ can be considered constant in it. If the actual environment meets the requirements only partly the raw data must be corrected for this, and $|\psi|^2\rmd^3x$ refers to the corrected data. 

The wavepacket is thus like a cloud moving along while triggering thunder and lightening here and there. One might imagine that small fluctuations in the environment determine the actual place of the thunder and lightening, but this is not the question we are concerned with. 

The effect induced by the wavepacket need not be observed. In realism, if an effect occurs, it occurs independently of our observing it. And it may be any effect. It is only when we want to refer to the same physical situation as the Copenhagen interpretation that it must be a special observable effect, observable according to the criteria of the Copenhagen interpretation. A stable macroscopic record of the effect is required, but how to achieve this is not a specific problem of quantum mechanics and does not concern us here. 

The postulate of normalization of each wavepacket, $\int|\psi|^2\rmd^3x=1$, does no longer mean that the probability of the entire sample space is unity; rather it means the postulate that each wavepacket of a certain kind represents the same amount of some conserved quantity, the nature of which is left open (cf. point (6) below). 

The interpretation in the case of quantities other than position is analogous.
As we may say that the expression $|\psi(\bfitx,t)|^2\rmd^3x$ gives the probability that the wavepacket induces an effect at $\bfitx$ (`in $\rmd^3x$ about $\bfitx$'), or acts at $\bfitx$ in position space, we may say that if the wavepacket in some  `$A$' space is $\phi(a)$, the expression $|\phi(a)|^2\rmd a$ gives the probability that the wavepacket acts at $a$ in $A$ space. This also applies to discrete eigenvalues where `at $a$' then means `at the point $a$' rather than `in d$a$ about $a$'. 

Thus neither a single, sharp position in ordinary space nor one in momentum space is a property of an elementary particle; only the ranges $\Delta x$ and $\Delta p_x$ are. The Heisenberg relations $\Delta x \Delta p_x \ge \hbar/2$, and the analogous relations between the ranges of other quantities, then express a relation between properties of wavepackets. They have nothing to do with any measurement. The importance of the Heisenberg relations stems from the importance of the wavepacket nature of the elementary particle. In this view the Heisenberg relations or, equivalently, the canonical commutation relations $[X,P_x] = \rmi \hbar$ are not basic structural features of the formalism and the procedure of canonical field quantization, namely to impose these relations on field quantities, loses its strongest support.
\vspace{6pt}

(4) A wavepacket has no internal structure. The wavepacket is an elementary region of space and must not be subdivided into smaller parts. 

This \emph{internal structurelessness} implies that the wavepacket in a sense behaves like a rigid body of Newtonian mechanics: imagine a large cigar-shaped wavepacket which interacts with another, small wavepacket at one end. The large (as well as the small) wavepacket is modified by this interaction, but the modification must not be imagined as starting from the one end and gradually expanding over the whole packet; rather the packet is simultaneously (in some Lorentz system) modified as a whole. The modification as such need not occur instantaneously but may take some time. In a second interaction with another small packet at the other end the large packet is already modified (at least to some extent) and the characteristics of the second interaction are thus influenced by the first. The time between the two interactions may be shorter than the time for a light signal to travel from one end to the other, so that the two correlated interactions occur with a spacelike separation in spacetime. 

It thus appears as if a signal had propagated inside the wavepacket with superluminal velocity. Dirac [1938] had already conceived such a possibility. He supposed the electron (though not its charge) to have a finite size, of the order of the classical electron radius $e^2/mc^2$, and he wrote:
\begin{quote}
it is possible for a signal to be transmitted faster than light through the interior of an electron. The finite size of the electron now reappears in a new sense, the interior of the electron being a region of failure, not of the field equations of electromagnetic theory, but of some of the elementary properties of space-time.
\end{quote}

In fact I would consider the transmission of effects inside the wavepacket as occurring not only with superluminal but with infinite velocity. However, I would propose a different viewpoint: it is not that the velocity is infinite, rather that there is no distance to travel, as there are no points to be distinguished inside the packet. And as there are no parts of a wavepacket there can be no interaction between them. Our interpretation thus is in line with the old conjecture (Lorentz [1909]; Feynman [1966]; Rohrlich [1973]) that there is no self-interaction, neither Coulomb nor radiative, nor of any other kind, between spatial parts of one and the same particle. We note, however, that this does not exclude the existence of recoil effects of the emitted radiation on the emitting wavepacket as a whole, as it is considered by Barut and collaborators in their approach to quantum electrodynamics without canonical quantization (Barut [1988b]; Barut and Dowling [1989]; and literature quoted there). 

In an interaction of the wavepacket with another one the whole of its charge, magnetic moment, etc. is involved. This accounts for the point-like behaviour of the wavepacket in the interaction. The Copenhagen school interprets this by assuming that besides the wave function there is a particle, a point particle, and this point carries all the quantities with it. This avoids the internal structurelessness but only at the price of the `wave-particle duality' problem. 

On the other hand, there are quantities like momentum or energy (of a free wavepacket) which have no single fixed value, and the wavepacket may act with different (though not quite sharp) values of them in different interactions. Nevertheless, in no case does this mean that only a spatial part of the packet were involved in the interaction. In point (6) below we shall extend this in that a coalesced wavepacket may act with a rational fraction of some of its quantities. 

The absence of structure comes into play only when the point-like intrinsic features of the interaction as such are considered, not when the probability of its occurrence is calculated. In this calculation the full $\psi(\bfitx,t)$ is used, defined for every point in the wavepacket. The wavepacket in a sense appears structured when seen from outside but unstructured when seen from inside. This is why I speak of internal structurelessness and not just of structurelessness. I see no other way to cope with the observed reality than first to introduce $\psi(\bfitx,t)$ for all values of $\bfitx$ and $t$ and then to restrict the meaning of $\bfitx$ and $t$ inside the wavepacket through additional postulates. In any case some postulate of this kind is inevitable if one wants to deal with objects that are extended but that are also fundamental and indivisible. The internal structurelessness might appear a most peculiar feature of the proposed interpretation. On the other hand, there are quite peculiar features of physical reality, for example those revealed in the Einstein-Podolsky-Rosen experiments (Section 5), and we shall see that internal structurelessness can give them a consistent interpretation.

\bigskip
\noindent
{\textbf{3~~Postulates: systems of particles}}
\smallskip

\noindent
Systems of identical (which we call similar) and systems of non-identical (dissimilar) particles are fundamentally different because in our interpretation similar particles can coalesce and form one single wavepacket but dissimilar particles cannot.
\vspace{6pt}

(5) In quantum mechanics systems of similar particles are described by symmetrized wave functions in configuration space $\Psi_{\textrm{\scriptsize{SA}}}(\bfitx_1,\bfitx_2,\ldots,\bfitx_N,t)$, that is, $\Psi_{\textrm{\scriptsize{SA}}}$ must be either symmetric or antisymmetric under any permutation of the particle labels $\bfitx_i$, which here are to include the spin labels. This expresses the fact that for the observer the particles are indistinguishable and lose their individuality. In our interpretation this loss of individuality is conceived to be an objective physical process: the original similar wavepackets coalesce and form one single wavepacket. Think of two water drops that coalesce into one. The coalesced wavepacket is a wavepacket in the sense of our interpretation, in particular with the property of internal structurelessness, and it may again coalesce with other packets. A necessary condition for coalescence to set in is that the wavepackets overlap, that is,
\[
\psi_1(\bfitx,t)\,\psi_2(\bfitx,t)\ldots\psi_N(\bfitx,t)\neq 0.
\]
Thus, symmetrization is essentially a recipe for mathematically coping with the physical process of coalescence; and though the coalesced wavepacket is described mathematically in multidimensional configuration space by means of the function $\Psi_{\textrm{\scriptsize{SA}}}$, physically it is to be conceived as an object in ordinary $(3 + 1)$-dimensional space. Also, we note that the symmetrized wave function $\Psi_{\textrm{\scriptsize{SA}}}$, does not in any case mean a coalesced packet. Consider a system of two electrons, one on the earth and the other on the moon, which have never overlapped and therefore cannot form a coalesced wavepacket. Yet, according to the general symmetrization postulate in standard quantum mechanics, the system has to be described by an antisymmetric wave function. On the other hand, in this case, as in all cases without overlap, symmetrization is innocuous because it does not lead to any effect that could not already be obtained without symmetrization (Pauli [1933]; Messiah [1961]).

There is coalescence both between similar Bose packets and between similar Fermi packets. The difference is that Bose packets can condense, but Fermi packets cannot. Condensation is the special case of coalescence where the coalesced wavepacket is describable by a wave function of the special product form $\phi(\bfitx_1,t)$ $\phi(\bfitx_2,t)\ldots\phi(\bfitx_N,t)$, with the same $\phi$ in all factors. This is another formulation of the Pauli exclusion principle.
\vspace{6pt}

(6) Any wavepacket represents an integer number of quanta. When $N$ similar wavepackets, which each represent one quantum, coalesce the coalesced packet represents $N$ quanta. In somewhat pictorial language one may say that a wavepacket contains an integer number of quanta, but not in the sense of a bag that contains balls but of a bag that contains water in integer multiples of some standard portion. The nature of the quanta is left open. 

In extension of postulate (4), which referred to an elementary (one-quantum) wavepacket, an $N$-quantum coalesced wavepacket may act with any number $(\geq 1)$ of its quanta in an interaction with another packet. Considering the quantization of charge, spin, etc. as special manifestations of the existence of quanta, we conceive that a coalesced packet may act with a fraction of its charge, etc., where, however, that fraction corresponds to an integer multiple of the elementary charge, etc. For example, a coalesced packet of two elementary proton packets has the charge $2e$, but it may act with the charge $1e$ only. 

This postulate distinguishes our interpretation from the neoclassical radiation theory of Jaynes and co-workers (Jaynes [1973]; Mandel [1976]) who try to get along without canonical quantization of the radiation field and without any substitute for it. In this neoclassical theory an electromagnetic pulse (wavepacket) that contains only one photon (quantum) can trigger two counters simultaneously, which is contrary to observation (Clauser [1974]; Grangier \emph{et al.} [1986]). According to our postulate a photon wavepacket that represents only one quantum can act only once, that is, in one counter only. In this respect the postulate resembles Einstein's way of explaining the photolectric effect. On the other hand, a wavepacket that contains $N$ photons in our interpretation may undergo up to $N$ interactions simultaneously (Hanbury Brown and Twiss effect, nonlinear optics).
\vspace{6pt}

(7) In addition to coalescence there is the inverse process of splitting of a coalesced wavepacket into several distinct packets, induced by some kind of interaction with another wavepacket around. When a coalesced $N$-quantum packet splits, the sum of all quanta represented by the final packets is also $N$. In other words, the number of quanta is conserved in coalescence and splitting. 

The postulates on condensation (5), integer number of quanta (6) and splitting (7) easily yield the Bose- and Fermi distributions by means of Einstein's method of balance relations (Jabs [1996]).
\vspace{6pt}

(8) Now we consider dissimilar particles. In the Copenhagen interpretation, where the wave function is the mathematical description, not of a physical object, but merely of the observed behaviour of a physical object, it does not appear to make much difference whether the object concerned is an electron, an atom, a macroscopic body, the universe, or even a fictitious point like the centre of mass. This is quite different in our interpretation where the wave function is the mathematical description or representation of the individual physical object itself. And the physical object that is so represented is only a wavepacket, that is, an elementary particle (mathematically represented by $\psi$), or a system of similar coalesced elementary particles (mathematically represented by $\Psi_{\textrm{\scriptsize{SA}}}$). 

A system of dissimilar elementary particles like an atom, a macroscopic body, or the universe, is not a wavepacket but an aggregate of wavepackets.
Generally, an aggregate is a set of distinct wavepackets, which may or may not dynamically interact with one another. The hydrogen atom, for example, is an aggregate of a proton wavepacket and an electron wavepacket. A system of two electrons, on the other hand, may be a wavepacket or an aggregate, depending on whether the two electron wavepackets have coalesced or not. In our conception of an aggregate we essentially replace the Newtonian mass points of the classical description by wavepackets. If all wavepackets of an aggregate were mathematically specified at all times the description would be complete; much as the classical description is complete if the positions and momenta of all the mass points are specified. 

The collapse or reduction process (Section 4) does not apply to aggregrates, only to wavepackets. There is thus no reduction of the aggregate as a whole, only the individual constituent wavepackets may be reduced. This avoids the problem of Schr\"odinger's cat (Jabs [1996]). 

The non-symmetrized configuration-space wave function $\Psi(\bfitx_1,\bfitx_2,\ldots,\bfitx_N,t)$ in our interpretation is merely a convenient calculational device that serves to compute some properties of aggregates, comparable, for example, to the partition function in statistical mechanics. This does not keep us from maintaining a realistic interpretation even in this situation: in analogy to point (3) above, the wavepackets of the aggregate have their properties even when not observed, and the mathematical expression $|\Psi(\bfitx_1,\bfitx_2,\ldots,\bfitx_N,t)|^2\rmd^3x_1\rmd^3x_2\ldots\rmd^3x_N$, for example, is interpreted as the probability that wavepacket 1 induces an effect at $\bfitx_1$ and wavepacket 2 at $\bfitx_2$, and so on. 

It would be reassuring if the concept of an aggregate as an assembly of wavepackets could be substantiated by a mathematical description in terms of ordinary $(3 + 1)$-dimensional wave functions, one for each wavepacket, without the use of any configuration-space wave function. In fact there exist two such descriptions: in canonical quantum field theory and in Feynman's space-time approach to quantum electrodynamics. However, use is there made of the procedure of canonical quantization, or equivalent procedures, which are rejected in our interpretation (point (3)). In standard nonrelativistic quantum mechanics an ordinary-space description of aggregates does not exist. This was the most serious objection to Schr\"odinger's original wavepacket picture. It can, however, not be mended without changing the formalism, and this is beyond the intended scope of the present work. In view of the fact that there exist already two ordinary-space descriptions, though unacceptable because of other features, it does not appear unreasonable to suppose that a satisfactory extension of the formalism will be possible. In particular I would like to point out to the work of Barut and collaborators in this direction (Barut [1988a, b, c]; Barut and Dowling [1989]; and literature cited there).

 This concludes the outline of the basic features. We will now discuss their role in the measurement and in the Einstein-Podolsky-Rosen problem.

\bigskip
\noindent
{\textbf{4~~The measurement problem}}
\smallskip

\noindent
In the Copenhagen interpretation there are quantities that come into existence only in the act of measurement, and it is only immediately after the (appropriate) measurement that the quantity is said to have a definite value, namely the measured one. 

In a realist interpretation a measurement measures what already exists. This implies that some operations which in the Copenhagen interpretation are called measurements in this interpretation are not. Two examples, one for continuous and one for discrete eigenvalues, may explain the general features. 

What already exists in the case of an electron that goes perpendicularly through a thin layer of photographic emulsion is the electron's wave function, or its width, for example. This width can be measured, but it is essential that we have an ensemble of equal electrons. We may let them pass through the emulsion layer, one after the other, and the distribution of the spots produced in the layer then images the width of the incoming electron wavepacket. What the electron wavepacket does after it has produced the spot is irrelevant for the concept of measurement in our interpretation. 

What already exists in the case of the atomic electron that enters a Stern-Gerlach apparatus is the direction of the electron's spin, in the sense that the incoming electron wavepacket is an eigenpacket of the spin-component operator with respect to some direction. This direction can be measured if we have an ensemble of electrons, all with the same spin direction, and let them pass through the Stern-Gerlach apparatus. In this physical situation the incoming wavepackets are forced to become eigenpackets of the spin-component operator with respect to the spin-reference axis of the apparatus. By observing the abundance ratios of up and down deflections for various orientations of the apparatus axis, the spin direction of the incoming electron can be calculated according to the standard formulas of quantum mechanics for spin-\mbox{${1\over 2}$}~particles. 

The most important point of the measurement problem is this. According to the Copenhagen quantum mechanics a wave function may vary in time in two ways: sometimes deterministically, as determined by the Schr\"odinger equation, and at other times with a random element in its behaviour, in the collapse or reduction process in a measurement. The problem is: what feature of the measurement makes the measurement interaction so different from the other one? 

In a realistic interpretation the reduction process is an objective physical process, independent of whether we observe it or not. And it does not contradict any experimental fact to assume that it is not instantaneous but takes a small time interval. Nor is the process, in our interpretation, restricted to situations which in the Copenhagen interpretation are called measurements. The problem is thus no longer a `measurement problem', but it remains a problem, and a serious one. It only assumes a different form: what feature of the physical environment makes the reduction process, rather than the deterministic evolution, occur? It is now the problem of how physically (and then mathematically) to characterise the reduction process in terms of current densities, field strengths, energy thresholds, etc. It clearly requires another extension of the current formalism. In any case, in a realistic interpretation the problem is made amenable to the methods of theoretical and experimental physics, rather than only to those of philosophy or psychology. See, for example, the experiments proposed or performed by Bussey [1984], [1986], [1987a]; Badurek \emph{et al.} [1986]; and Summhammer \emph{et al.} [1987].

\bigskip
\noindent
{\textbf{5~~The Einstein-Podolsky-Rosen problem}}
\smallskip

\noindent
The essential features of the Einstein-Podolsky-Rosen (EPR) problem (Einstein et al.[1935]; Bohm [1951]) may be described in the following way: two protons, 1 and 2, which have interacted in the past form a system of zero total spin and fly apart with equal velocities in opposite directions. Proton 1 then enters a Stern-Gerlach-type apparatus A, and proton 2 another apparatus B. Each proton in its respective apparatus turns into a pure spin-up or a pure spin-down proton with respect to the axis of its apparatus. Let us assume that the distance between the place 0 of interaction and the apparatus A is shorter than that between 0 and apparatus B, so that proton 1 enters apparatus A before proton 2 enters B; and let us assume that proton 1 in A turns into a spin-up proton with respect to the axis of A. Then, at the same instant proton 2 turns into a spin-down proton with respect to the axis of A although the distance between proton 2 and apparatus A may be arbitrarily large. Since the axis of A can be adjusted immediately before the arrival of proton 1 there is some kind of simultaneous action-at-a-distance. In speaking of separated protons with each having its own definite spin axis, one is using a language which has to be used warily. If one insists on a description purely in terms of outcomes of experiments it is not so easy to see that an action-at-a-distance or nonlocality is involved. According to quantum mechanics the joint probability that apparatus A obtains an $r_{\textrm{\scriptsize{A}}}$-proton ($r_{\textrm{\scriptsize{A}}}=+1$: spin up, 
$r_{\textrm{\scriptsize{A}}}=-1$: spin down) and apparatus B an
$r_{\textrm{\scriptsize{B}}}$-proton is $P_1(r_{\textrm{\scriptsize{A}}},r_{\textrm{\scriptsize{B}}}|\zeta_{\textrm{\scriptsize{ab}}})=\frac{1}{4}(1-r_{\textrm{\scriptsize{A}}}r_{\textrm{\scriptsize{B}}}\cos\zeta_{\textrm{\scriptsize{ab}}})$, where $\zeta_{\textrm{\scriptsize{ab}}}$ is the angle between the axes of the two apparatuses. Only the fact that this formula implies correlations that can violate the Bell inequality reveals its nonlocal character (Bell [1981]). Though this nonlocality cannot be employed for sending superluminal messages (Jordan [1983]; Bussey [1987b]; and literature cited there) the situation remains striking enough. Yet the experiments which have been devised and carried out in order to test this particular
prediction of quantum mechanics have, within their limits, confirmed it (Lamehi-Rachti and Mittig [1976]; Clauser and Shimony [1978]; Aspect and Grangier [1985]; Perrie \emph{et al.} [1985]; Rarity and Tapster [1990]; and literature cited there). 

In the proposed interpretation the mechanism for inducing the above EPR correlations is coalescence of similar wavepackets and internal structurelessness: when the two proton wavepackets come together they coalesce. The coalesced packet then spreads out towards the apparatuses A and B. When it reaches apparatus A it interacts with this apparatus, so that the packet first is modified simultaneously over its whole extension and then splits into two independent packets which have their spin components correlated according to the formulas of quantum mechanics for similar particles. 

There is thus, in this interpretation, no `telepathy' between different objects but only within one and the same object. This is a weaker form of `telepathy' than in Copenhagen quantum mechanics. In Copenhagen quantum mechanics the two particles are interrelated until a reduction occurs. The reduction occurs only in an observation. The observation is at the observer's disposal, and he or she may perform it an arbitrarily long time after the interaction, when the particles are separated from each other by an arbitrarily large distance. In our interpretation the range of the nonlocality is limited by the extension of the coalesced wavepacket, and this, in realism, is an objective physical criterion. Moreover, there are the objective processes of splitting and of reduction, induced by other wavepackets in the surroundings of the considered wavepacket, and due to these processes the wavepacket may have only little chance to spread out over a very large region. 

Whether a wavepacket, coalesced or elementary, in order to maintain internal structurelessness must be spatially connected (simply or not) or whether it may consist of non-overlapping parts is another question. Of course, it would be nice if it could always be considered as connected. The question can, in principle, be answered by experiments on EPR correlations and reduction processes, but in practice it is not so simple because at the present stage it is not clear when exactly two parts are to be considered separated, It happens that most wave functions mathematically extend to infinity so that there would always be nonzero overlap. Moreover, the degree of separation between two parts may be judged by comparing their mutual distance with the widths of the parts themselves, but the exact definition of these widths is arbitrary to a considerable degree. With some reasonable assumptions on these widths a closer examination of the existing EPR experiments seems to indicate that the coalesced packets involved might still be connected (Jabs [1996]). On the other hand, elementary wavepackets in a Stern-Gerlach magnet or in a neutron interferometer (Summhammer \emph{et al.} [1987]) seem to develop into really separated parts. Therefore we think that we must be prepared to accept that wavepackets can consist of disconnected parts, and that this applies to elementary and to coalesced packets, the two being of the same nature.

\bigskip
\noindent
{\textbf{6~~Crucial experiments}}
\smallskip

\noindent
The experiments we consider here are concerned with the concepts of coalescence, internal structurelessness and splitting.

(a) The first experiment is a variant of an already performed one. It is concerned with EPR correlations, as described in Section 5. These correlations have only been observed between similar particles, namely with pairs of photons and with pairs of protons. In these cases we predict the same results as the Copenhagen interpretation. However, since in our interpretation the EPR correlations are a consequence of coalescence and internal structurelessness of the coalesced wavepacket and since coalescence occurs only between similar wavepackets, we predict that EPR correlations will not occur between dissimilar (`non-identical') particles. Thus, if the proton-proton scattering experiment described in the preceding section, which has been carried out by Lamehi-Rachti and Mittig [1976], were to be repeated with the incoming proton replaced by a neutron or a positron, we predict no longer the EPR correlations implied by the formula $P_1$ of Section 5. We conjecture that only the weaker correlations, implied by the formula $P_2(r_{\textrm{\scriptsize{A}}},r_{\textrm{\scriptsize{B}}}|\zeta_{\textrm{\scriptsize{ab}}})=\frac{1}{4}(1-\frac{1}{3}r_{\textrm{\scriptsize{A}}}r_{\textrm{\scriptsize{B}}}\cos\zeta_{\textrm{\scriptsize{ab}}})$, would be observed. $P_2$ follows from the assumption that after the interaction the two spin-$\frac{1}{2}$ particles have their spins in opposite directions but are completely separated from each other in every respect. $P_1$ can lead to a violation of the Bell inequality, but $P_2$ never can (Jabs [1996]). The Copenhagen interpretation prescribes the use of the same formula $P_1$ in both cases. 

There is one possible exception, namely pairs of particle and antiparticle.
These are dissimilar, and no symmetrization postulate applies. Nevertheless, in the mathematical treatment of electron-positron scattering in quantum electrodynamics virtual-annihilation terms appear, and due to the `substitution law' these terms have the same mathematical structure as exchange terms, that is, effects of symmetrization and hence coalescence (Bjorken and Drell [1964], Jauch and Rohrlich [1976]). We expect that this carries over to any particle-antiparticle pair, and therefore we anticipate that in such pairs coalescence and hence EPR correlations may occur.

(b) Even for systems of similar particles the Copenhagen and the realistic interpretation here presented differ in some experimental predictions. For example, the wavepacket $\Psi_{\textrm{\scriptsize{SA}}}(\bfitx_1,\bfitx_2,t)$, when it expands from the place O of interaction towards the apparatuses A and B, may interact with other wavepackets around. There may thus occur some unobserved spurious scattering events with the effect that
$\Psi_{\textrm{\scriptsize{SA}}}$ is modified before it reaches A or B so
that the correlations between the results at A and at B are decreased. Now, in the realistic interpretation these scattering events may have the additional effect of splitting the packet $\Psi_{\textrm{\scriptsize{SA}}}$, and this leads to an additional decrease in the correlations. In the Copenhagen interpretation this splitting does not occur since here the splitting is connected with a reduction, and the reduction occurs only in the measurement in apparatus A or B. The decrease in the correlations is thus stronger in the realistic than in the Copenhagen interpretation. 

One could study this decrease by putting a variable amount of material between the place O of interaction and the apparatus A (or B) or by varying the distance OA in the case that there is `vacuum' (i.e.~a rarefied gas) between O and A. Unlike the experiment in (a), this type of experiment does not allow a quantitative prediction since the details of the splitting interactions are unknown at present. Reversely, one could use the observed deviations from the predictions of the Copenhagen quantum mechanics in this situation to specify the splitting interactions. Experiments with variation of the distance OA have been performed with photon pairs from $e^+e^-$ annihilation. Faraci \emph{et al.} [1974] reported a change in the correlations, whereas the later experiments (Wilson \emph{et al.} [1976]; Bruno \emph{et al.} [1977]) did not detect any significant change. One might, however, suspect that the considered distances of less than 2.5 m were not large enough. Experiments with a variable amount of material interposed between O and A have also been proposed by Jauch ([1971], p.~38). To my knowledge they have not been performed so far.
\emph{
\begin{flushright}
Sektion Physik, Theoretische Physik \\Theresienstr. 37, Universit\"at M\"unchen, Germany \\(On leave of absence from \\Department of Physics\\ Federal University of Para\'{\i}ba \\Jo\~{a}o Pessoa, Brazil)
\end{flushright}}

\bigskip
\noindent
{\textbf{References}}
\begin{description}
\renewcommand{\labelenumi}{[\arabic{enumi}]}
\begin{sloppypar}

\item ASPECT, A. and GRANGIER, P. [1985]: About Resonant Scattering and Other Hypothetical Effects in the Orsay Atomic-Cascade Experiment Test of Bell Inequalities: A Discussion and Some New Experimental Data. Lett. Nuovo Cimento \textbf{43}, pp. 345-8.

\item BADUREK, G., RAUCH, H. and TUPPINGER, D. [1986]: Neutron Interferometric Double-Resonance Experiment. Phys. Rev. A \textbf{34}, pp. 2600-8. 

\item BALLENTINE, L. E. [1970]: The Statistical Interpretation of Quantum Mechanics. Rev. Mod. Phys. \textbf{42}, pp. 358-81.

\item BARUT, A. 0. [1988a]: Schr\"odinger's Interpretation of $\psi$ as a Continuous Charge Distribution. Annalen der Physik \textbf{45}, pp. 31-6.

\item BARUT, A. 0. [1988b]: Quantum-Electrodynamics Based on Self-Energy. Physica Scripta \textbf{T21}, pp. 18-21.

\item BARUT, A. 0. [1988c]: The Revival of Schr\"odinger's Interpretation of Quantum Mechanics. Found. Phys. Lett. \textbf{1}, pp. 47-56.

\item BARUT, A. 0. and DOWLING, P. [1989]: Quantum Electrodynamics Based on Self-fields, Without Second Quantization: Apparatus-Dependent Contributions to $g$-$2$. Phys. Rev. A \textbf{39}, pp. 2796-805.

\item BELL, J. S. [1973]: Subject and Object. in: Mehra (ed.) [1973], pp. 687-90.

\item BELL, J. S. [1981]: Bertlmann's Socks and the Nature of Reality. Journal de Physique \textbf{42}, pp. C2-41-62. Reproduced in: Speculations in Science and Technology [1987], \textbf{10}, pp. 269-85, and in: Speakable and Unspeakable in Quantum Mechanics, pp. 139-58. Cambridge: Cambridge University Press [1987].

\item BJORKEN, J. D. and DRELL, S. D. [1964]: Relativistic Quantum Mechanics, section 7.9. New York: McGraw-Hill.

\item BOHM, D. [1951]: Quantum Theory, pp. 614-22. New York: Prentice-Hall.

\item BOHM. D., HILEY, B. J. and KALOYEROU, P. N. [1987]: An Ontological Basis for the Quantum Theory. Phys. Reports \textbf{144}. pp. 321-75.

\item BRUNO, M., D'AGOSTINO, M. and MARONI, C. [1977]: Measurement of Linear Polarization of Positron Annihilation Photons. Nuovo Cimento \textbf{40B}, pp. 143-52.

\item BUNGE, M. [1977]: Quantum Mechanics and Measurement. Int. J. Quant. Chem. \textbf{12}, pp. 1-13.

\item BUNGE, M. and KALNAY, A. J. [1975]: Welches sind die Besonderheiten der Quantenphysik gegenüber der klassischen Physik? in: R. Haller and J. G\"otschl (eds.): Philosophie und Physik, pp. 25-38. Braunschweig: Vieweg. 

\item BURGOS, M. E. [1984]: An Objective Interpretation of Orthodox Quantum Mechanics. Found. Phys. \textbf{14}, pp. 739-52.

\item BUSSEY, P. J. [1984]: When Does the Wavefunction Collapse? Phys. Lett. \textbf{106A}, pp. 407-9.

\item BUSSEY, P. J. [1986]: Wavefunction Collapse and the Optical Theorem. Phys. Lett. A \textbf{118}, pp. 377-80. 

\item BUSSEY, P. J. [1987a]: The Fate of Schr\"odinger's Cat. Phys. Lett. A \textbf{120}, pp. 51-3.

\item BUSSEY, P. J. [1987b]: Communication and Non-Communication in Einstein-Rosen Experiments. Phys. Lett. A \textbf{123}, pp. 1-3.

\item CLAUSER, J. F. [1974]: Experimental Distinction Between the Quantum and Classical Field-theoretic Predictions for the Photoelectric Effect. Phys. Rev. D \textbf{9}, pp. 853-60.

\item CLAUSER, J. F. and SHIMONY, A. [1978]: Bell's Theorem: Experimental Tests and Implications. Rep. Prog. Phys. \textbf{41} pp. 1881-1927.

\item DIRAC, P. A. M. [1938]: Classical Theory of Radiating Electrons. Proc. R. Soc. Lond. A \textbf{167}, pp. 148-69.

\item EINSTEIN, A., PODOLSKY, B. and ROSEN, N. [1935]: Can Quantum-Mechanical Description of Physical Reality Be Considered Complete? Phys. Rev. \textbf{47}, pp. 777-80.

\item FARACI, G., GUTKOWSKI, D., NOTARRIGO, S. and PENNISI, A. R. [1974]: An Experimental Test of the EPR Paradox. Lett. Nuovo Cimento \textbf{9}, pp. 607-11.

\item FEYNMAN, R. P. [1966]: The Development of the Space-Time View of Quantum Electrodynamics. Physics Today \textbf{19}, 8, pp. 31-44.

\item GRANGIER, P., ROGER, G. and ASPECT, A. [1986]: Experimental Evidence for a Photon Anticorrelation Effect on a Beam Splitter: A New Light on Single-Photon Interferences. Europhys. Lett. \textbf{1}, pp. 173-9.

\item HEISENBERG, W. [1958]: The Representation of Nature in Contemporary Physics. Daedalus \textbf{87}, pp. 95-108.
 
\item JABS, A. [1996]: Quantum Mechanics in Terms of Realism. arXiv:quant-ph/9606017.
 
\item JAUCH, J. M. [1971]: Foundations of Quantum Mechanics. in: B. D'Espagnat (ed.): Foundations of Quantum Mechanics (Proc. of the Intern. School of Physics `Enrico Fermi', Course 40), pp. 20-55. New York: Academic. 

\item JAUCH, J. M. and ROHRLICH, F. [1976]: The Theory of Photons and Electrons, section 12-2. New York: Springer.

\item JAYNES, E. T. [1973]: Survey of the Present Status of Neoclassical Radiation Theory. in: L. Mandel and E. Wolf (eds.): Coherence and Quantum Optics. pp. 35-81. New York: Plenum.
 
\item JORDAN, T. F. [1983]: Quantum Correlations Do Not Transmit Signals. Phys. Lett. \textbf{94A}, p. 264.

\item LAMEHI-RACHTI, M. and MITTIG, W. [1976]: Quantum Mechanics and Hidden Variables: A Test of Bell's Inequality by the Measurement of the Spin Correlation in Low-energy Proton-proton Scattering. Phys. Rev. D \textbf{14}, pp. 2543-55.

\item L\'EVY-LEBLOND, J.-M. [1976]: Towards a Proper Quantum Theory. Dialectica \textbf{30}, pp. 161-96. 

\item LORENTZ, H. A. [1909]: The Theory of Electrons. Leipzig: Teubner, 2nd edn. [1952], p. 215. New York: Dover.

\item MANDEL, L. [1976]: The Case for and against Semiclassical Radiation Theory. in: E. Wolf (ed.): Progress in Optics XIII, pp. 27-68. Amsterdam: North-Holland. 

\item MAXWELL, N. [1982]: Instead of Particles and Fields: A Micro Realistic Quantum ``Smearon'' Theory. Found. Phys. \textbf{12}, pp. 607-31.

\item MEHRA, J. (ed.) [1973]: The Physicist's Conception of Nature. Dordrecht: Reidel.

\item MESSIAH, A. [1961]: Quantum Mechanics, Vol. II, §XIV.8. Amsterdam: North-Holland.

\item PAULI, W. [1933]: Die allgemeinen Prinzipien der Wellenmechanik. in: H. Geiger and K. Scheel (eds.): Handbuch der Physik, Vol. 24. Part I, pp. 83-272, esp. pp. 192, 193. Berlin: Springer (English translation: General Principles of Quantum Mechanics [1980]. Berlin: Springer).

\item PERRIE, W., DUNCAN, A. J., BEYER, H. J. and KLEINPOPPEN, H. [1985]: Polarization Correlation of Two Photons Emitted by Metastable Atomic Deuterium: A Test of Bell's Inequality. Phys. Rev. Lett. \textbf{54}, pp. 1790-3.

\item POPPER, K. R. [1985]: Realism in Quantum Mechanics and a New Version of the EPR Experiment. in: G. Tarozzi and A. van der Merwe (eds.): Open Questions in Quantum Physics, pp. 3-25. Dordrecht: Reidel. 

\item RARITY, J. G. and TAPSTER, P. R. [1990]: Experimental Violation of Bell's Inequality Based on Phase and Momentum. Phys. Rev. Lett. \textbf{64}, pp. 2495-8.

\item RAYSKI, J. [1973]: The Possibility of a More Realistic Interpretation of Quantum Mechanics. Found. Phys. \textbf{3}, pp. 89-100.

\item ROBERTS, K. V [1978]: An Objective Interpretation of Lagrangian Quantum Mechanics. Proc. R. Soc. Lond. A \textbf{360}, pp. 135-60.

\item ROHRLICH, F. [1973]: The Electron: Development of the First Elementary Particle Theory, in: Mehra [1973], pp. 331-69.

\item ROHRLICH, F. [1987]: Reality and Quantum Mechanics. Annals of the New York Academy of Sciences \textbf{480}, pp. 373-81.

\item SCHR\"ODINGER, E. [1926]: Der stetige \"Ubergang von der Mikro- zur Makromechanik. Naturwiss. \textbf{14}, pp. 664-6 (English translation in: Collected Papers on Wave Mechanics [1982], pp. 41-4. New York: Chelsea).

\item SCHR\"ODINGER, E. [1928]: Four Lectures on Wave Mechanics. Reprinted in Co11ected Papers on Wave Mechanics [1982], p.~170 \emph{passim}. New York: Chelsea. 

\item STAPP, H. F. [1985]: Bell's Theorem and the Foundations of Quantum Physics. Am. J. Phys. \textbf{53}, pp. 306-17.

\item SUMMHAMMER, J., RAUCH, H. and TUPPINGER, D. [1987]: Stochastic and Deterministic Absorption in Neutron-interference Experiments. Phys. Rev. A \textbf{36}, pp. 4447-55.

\item WILSON, A. R., LOWE, J. and BUTT, D. K. [1976]: Measurement of the Relative Planes of Polarization of Annihilation Quanta as a Function of Separation Distance. J. Phys. G \textbf{2}, pp. 613-23.

\item WITTGENSTEIN, L. [1953]: Philosophical Investigations, \S402. Oxford: Basil Blackwell.

\item WOHL, C. G. \emph{et al.} [1984]: Review of Particle Properties. Rev. Mod. Phys. \textbf{56}, 2, Part II.
\end{sloppypar}
\end{description}
\vspace{-0.1cm}
\hspace{5cm}
-----------------------------

\medskip
\noindent
{\textbf{Note added when posting the paper on the arXiv}}
\vspace{6pt}

\noindent
The original text is left unaltered, except that a few typographical errors have been corrected and the preprint version Jabs [1991] has been replaced by the published version Jabs~[1996].

Several conceptions in the paper have changed since 1992. The most important changes are:
(1) Both similar and dissimilar particles can `coalesce', i.e., get entangled, and the term `coalesced' should everywhere be replaced by `entangled'. (2) The interpretation of  $|\psi(\bfitx,t)|^2\rmd^3x$ is no longer based on transition-probability formulas but on a physical criterion for the reduction process (arXiv:1204.0614). (3) The `splitting' introduced in Section 3(7) is identified with the reduction process. (4) The experiments described in Section 6 are no longer considered as crucial though still as worth being performed.
\medskip
\\Arthur Jabs\\
Alumnus, Technische Universit\"at Berlin.
\\
Vo\ss str.~9, 10117 Berlin, Germany
\\arthur.jabs@alumni.tu-berlin.de
\\(7 September 2014)
 
\end{document}